\documentclass[aps,preprint,showpacs,showkeys]{revtex4}
\usepackage{mathrsfs}
\usepackage{graphicx}
\usepackage{subfigure}
\usepackage{amssymb}
\usepackage{amsmath}
\usepackage{bm}
\usepackage{latexsym}
\usepackage{hyperref}

\begin{document}

\date{\today }
\title{The dynamics of a qubit in a spin-boson environment: a comparison between analytical and numerical method}
\author{Peihao Huang$^{1,2}$}
\email{phhuang@sjtu.edu.cn}
\author{Hang Zheng$^{1}$}
\author{Keiichiro Nasu$^{2}$}
\affiliation{$^1$Department of Physics, Shanghai Jiao Tong University, Shanghai 200240,
P.R.China\\
$^2$Solid State Theory Division, Institute of Materials Structure Science, High Energy Accelerator Research Organization (KEK), Tsukuba, Ibaraki 305-0801, Japan}

\begin{abstract}
The dynamics of a qubit under the decoherence of a two level fluctuator
(TLF) in addition to its coupling to a bosonic bath is investigated
theoretically. Two different methods are applied and compared for this
problem. One is a perturbation method based on a unitary transformation.
With the merit of our unitary transformation, non-adiabatic effect can be
taken into account efficiently. And the other one is the numerically exact
method, namely the quasi-adiabatic path-integral (QUAPI) propagator
technique. We find that the analytical method works well for a wide
parameter range and show good agreement with QUAPI. On the other hand, The
enhancement and the reduction of quantum decoherence of the qubit is checked
with varying bath temperature $T$ and TLF-bath coupling.

\end{abstract}

\maketitle


\section{Introduction}

Dissipative quantum dynamics is one of the central paradigm in the
theoretical physics because of its relation to the various physical and
chemical phenomena from the spontaneous emission to electron transfer in
molecular, from qubit decoherence to photon harvest in photosynthesis \cite%
{Nakamura1999,Chiorescu2004,You2005,Engel2007,Lee2007}. In the last decades,
spin-boson model, the simplest possible model to describe dissipation, is
studied intensively and offers a comprehensive understanding of the
decoherence phenomenon.

In this paper, a variant of the spin-boson model is studied, the overall
spin-boson model act as an environment of another spin. This model describes
a qubit in dissipative two-level fluctuators (TLF) which is believed to be
relevant to the prevailing 1/f noise in the Josephson qubits\cite%
{Paladino2002,Simmonds2004,Falci2005,Shnirman2005,Abel2008,Neeley2008,Lupascu2009,Lisenfeld2010}%
. Only a single spin-boson environment is considered since the qubit
dynamics is usually dominated by a particular TLF near resonance with the
qubit. An accurate evaluation of this problem is challenging. By employing
the flow equation method, Gassmann et.al. studied the spectrum of the
correlation functions of this model. And it is studied a lot recently with
various perturbation approaches \cite%
{Gassmann2002,Gassmann2005,Paladino2006,Paladino2008,Oxtoby2009,Lupascu2009,Lisenfeld2010}%
.

In this paper we examine the effect of the bath temperature and system bath
coupling on the qubit dynamics. One commonly believed concept is that
temperature and the coupling to noisy bath only play a negative role in
preserving the qubit coherence. However, it is pointed out in Ref. \cite%
{Montina2008} that the temperature can help the coherence when the qubit is
coupled to a TLF (or spin-boson) environment. To examine this effect we
treat the model more rigorously by consider spin-spin as a central quantum
system which coupled to a bosonic bath and compared with the result of the numerically exact
method, namely the quasi-adiabatic path-integral (QUAPI) propagator
technique. Good agreement is achieve between these two methods. And we find that it is possible for the reduction of decoherence with increasing temperature or with increasing TL-bath coupling, which verifies the previous findings.


The model is given as ($\hbar =1$): $H=H_{A}+H_{AB}+H_{B}$ with
\begin{eqnarray}
H_{A} &=&\frac{\Delta _{A}}{2}\sigma _{z}^{A},\,\,\,\,\,\,\,\,H_{AB}=\frac{%
g_{0}}{2}\sigma _{x}^{A}\sigma _{x}^{B},  \notag \\
H_{B+bath} &=&\frac{\Delta _{B}}{2}\sigma _{z}^{B}+\sum_{k}{\omega _{k}}%
b_{k}^{\dagger }b_{k}+\frac{\sigma _{x}^{B}}{2}\sum_{k}g_{k}(b_{k}^{\dagger
}+b_{k}),  \label{Hamiltonian}
\end{eqnarray}%
where TSSs are characterized by pseudospin-1/2 operators $\sigma _{z}^{A}$
and $\sigma _{z}^{B}$ as usual, $b_{k}$ and $b_{k}^{\dag }$ are the
annihilation and creation operators of the bath mode. $g_{0}$ and $g_{k}$
are the coupling constants. Here, the anisotropic coupling between TSS-A and
TSS-B subject to the z direction, and the B-bath coupling also to the z
direction. The bath is fully defined by the spectral density $J(\omega
)\equiv \sum_{k}g_{k}^{2}{\delta }(\omega -\omega _{k})$.
We will use the piezoelectric spectral density, which describes the
decoherence of a double quantum dots(DQD) qubit manufactured with GaAs \cite%
{Brandes1999,Wu2005},
\begin{equation}
J^{TL}(\omega )=\alpha \omega \left( 1-\frac{\omega _{d}}{\omega }\sin \frac{%
\omega }{\omega _{d}}\right) e^{-\omega ^{2}/2\omega _{l}^{2}},  \label{J_TL}
\end{equation}%
where $\omega_d$ is related to the center to center distance, and $\omega_l$ to the dot size.
Typically, $\omega_d\sim0.01(ps)^{-1}$ and $\omega_l\sim1(ps)^{-1}$\cite%
{Wu2005}. In the limit of $\omega_d\rightarrow0$, one can find that, Eq.~(\ref{J_TL}) goes back to the widely used Ohmic spectrum\cite{Leggett1987,Weiss1999}.

\section{Unitary transformation}
On the analogy with Ref.~\cite{Zheng2004}, we apply a unitary transformation
to the Hamiltonian, $H^{\prime }=\exp (S)H\exp (-S)$, with the generator $%
S\equiv \sum_{k}\frac{g_{k}}{2\omega _{k}}\xi _{k}(b_{k}^{\dag
}-b_{k})\sigma _{x}^{B}+i\theta _{0}\sigma _{y}^{A}\sigma _{x}^{B}$.
Therefore,%
\begin{eqnarray}
H_{A}^{\prime } &=&\frac{\sigma _{z}^{A}}{2}\left( \Delta _{A}\cos \theta
_{0}+g_{0}\sin \theta _{0}\right) \\
H_{AB}^{\prime } &=&\frac{1}{2}\sigma _{x}^{A}\sigma _{x}^{B}\left(
g_{0}\cos \theta _{0}-\Delta _{A}\sin \theta _{0}\right) -\frac{1}{2}\eta
\Delta _{B}\sin \theta _{0}i\sigma _{y}^{A}i\sigma _{y}^{B} \\
H_{B+bath}^{\prime (0)} &=&\frac{\sigma _{z}^{B}}{2}\eta \Delta _{B}\cos
\theta _{0}+\sum_{k}{\omega _{k}}b_{k}^{\dag }b_{k}+\sum_{k}\frac{g_{k}^{2}}{%
4\omega _{k}}(\xi _{k}^{2}-2\xi _{k}) \\
H_{B+bath}^{\prime (1)} &=&\frac{\sigma _{x}^{B}}{2}\sum_{k}g_{k}(1-\xi
_{k})(b_{k}^{\dagger }+b_{k})-\frac{\eta \Delta _{B}\cos \theta _{0}}{2}%
i\sigma _{y}^{B}X_{1} \\
H_{B+bath}^{\prime (2)} &=&\frac{\Delta _{B}\cos \theta _{0}}{2}\left[
\sigma _{z}^{B}\left( \cosh X_{1}-\eta \right) -i\sigma _{y}^{B}\left( \sinh
X_{1}-\eta X_{1}\right) \right]  \notag \\
&+&\frac{\Delta _{B}\sin \theta _{0}}{2}i\sigma _{y}^{A}\left[ \sigma
_{z}^{B}\sinh X_{1}-i\sigma _{y}^{B}\left( \cosh X_{1}-\eta \right) \right] ,
\end{eqnarray}%
where $X_{1}=\sum_{k}\frac{g_{k}}{\omega _{k}}\xi _{k}(b_{k}^{\dag }-b_{k})$
and $\eta $ is the thermodynamic average of $\cosh {X}_{1}$,
\begin{equation}
\eta =\exp \left[ -\sum\limits_{k}\frac{g_{k}^{2}}{2\omega _{k}^{2}}\xi
_{k}^{2}\coth (\beta \omega _{k}/2)\right] ,
\end{equation}%
which insures $H_{B+bath}^{\prime (2)}$ contains only the terms of two-boson
and multi-boson non-diagonal transitions and its contribution to physical
quantities is $(g_{k}^{2})^{2}$ and higher. Suppose the B-Bath coupling is
not strong, the last term of the above Hamiltonian are dicarded in the
following discussion. Now if we let%
\begin{eqnarray}
\Delta _{A}^{\prime } &=&\Delta _{A}\cos \theta _{0}+g_{0}\sin \theta
_{0},\Delta _{B}^{\prime }=\eta \Delta _{B}\cos \theta _{0}, \\
g_{0}^{\prime } &=&\left( g_{0}\cos \theta _{0}-\Delta _{A}\sin \theta
_{0}\right) =\eta \Delta _{B}\sin \theta _{0}, \\
g_{k}^{\prime } &=&g_{k}(1-\xi _{k})=\eta \Delta _{B}\cos \theta _{0}\frac{%
g_{k}}{\omega _{k}}\xi _{k},
\end{eqnarray}%
then,%
\begin{eqnarray}
&&\tan \theta _{0}=\frac{g_{0}}{\Delta _{A}+\eta \Delta _{B}}, \\
&&\xi _{k}=\frac{\omega _{k}}{\omega _{k}+\eta \Delta _{B}\cos \theta _{0}}.
\end{eqnarray}%
and the Hamiltonian becomes%
\begin{eqnarray}
H_{A,B}^{\prime (0)} &=&\frac{\Delta _{A}^{\prime }}{2}\sigma _{z}^{A}+\frac{%
\Delta _{B}^{\prime }}{2}\sigma _{z}^{B},\,\,\,\,H_{bath}^{\prime
(0)}=\sum_{k}{\omega _{k}}b_{k}^{\dagger }b_{k} \\
H_{AB}^{\prime } &=&g_{0}^{\prime }\left( \sigma _{+}^{A}\sigma
_{-}^{B}+\sigma _{-}^{A}\sigma _{+}^{B}\right) ,\,\,\,\,H_{B+bath}^{\prime
(1)}=\sum_{k}g_{k}^{\prime }\left( \sigma _{+}^{B}b_{k}+\sigma
_{-}^{B}b_{k}^{\dag }\right) .
\end{eqnarray}%
where we have omitted the constant and the second order terms.

It is easy to check that $\left( H_{AB}^{\prime }+H_{B+bath}^{\prime
(1)}\right) |g_{0}\rangle =0$, where $|g_{0}\rangle $ is the ground state of
$H_{A,B}^{\prime (0)}+H_{bath}^{\prime (0)}$. Therefore, the ground state
energy is $E_{g}=-\frac{1}{2}\Delta _{A}^{\prime }-\frac{1}{2}\Delta
_{B}^{\prime }-\sum\limits_{k}\frac{g_{k}^{2}}{4\omega _{k}}\xi _{k}(2-\xi
_{k})$. Note that, the transformed Hamiltonian $H^{\prime }=H_{A,B}^{\prime
(0)}+H_{bath}^{\prime (0)}+H_{AB}^{\prime }+H_{B+bath}^{\prime (1)}$ is of
similar form of that of RWA, which enable us to treat the system bath
coupling much easier, but $\Delta _{A}$, $\Delta _{B}$, $g_{0}/2$ and $%
g_{k}/2$ are replaced by $\Delta _{A}^{\prime }$, $\Delta _{B}^{\prime }$
and $g_{0}^{\prime }$ and $g_{k}^{\prime }$ due to the contributions of
anti-rotating terms. The renormalized effective coupling
become much smaller than the original coupling, which enable the pertubation
treatment works better than the ordinary Born approximation directly from
original Hamiltonian.

Now, we can diagonalize $H_{A,B}^{\prime (0)}+H_{AB}^{\prime }$ by a unitary
transformation $T$,%
\begin{equation}
T=\left[
\begin{array}{cccc}
1 & 0 & 0 & 0 \\
0 & \cos \left( \theta \right) & \sin \left( \theta \right) & 0 \\
0 & \sin \left( \theta \right) & -\cos \left( \theta \right) & 0 \\
0 & 0 & 0 & 1%
\end{array}%
\right] .
\end{equation}%
where $\tan 2\theta =2g_{0}^{\prime }/(\Delta _{A}^{\prime }-\Delta
_{B}^{\prime }).$ Then the Hamiltonian becomes $\widetilde{H}=\widetilde{H}%
_{0}+\widetilde{H}_{1}$,%
\begin{eqnarray}
\widetilde{H}_{0} &=&\sum_{i=0}^{3}\varepsilon _{i}|i\rangle \langle
i|+\sum_{k}{\omega _{k}}b_{k}^{\dagger }b_{k} \\
\widetilde{H}_{1} &=&\sum_{k}g_{k}^{\prime }\left( \varrho _{+}b_{k}+\varrho
_{-}b_{k}^{\dag }\right) .
\end{eqnarray}%
where, $\varepsilon _{3}=-\varepsilon _{0}=\frac{E_{p}}{2}\equiv \frac{1}{2}%
\left( \Delta _{A}^{\prime }+\Delta _{B}^{\prime }\right) $, $\varepsilon
_{2}=-\varepsilon _{1}=\frac{E_{m}}{2}\equiv \sqrt{\frac{1}{4}\left( \Delta
_{A}^{\prime }-\Delta _{B}^{\prime }\right) ^{2}+g_{0}^{\prime 2}}$, $%
\varrho _{+}=\varrho _{-}^{\dag }\equiv \cos \theta (|3\rangle \langle
2|-|1\rangle \langle 0|)+\sin \theta (|2\rangle \langle 0|+|3\rangle \langle
1|)$.

\section{Master Equation method}

Now, we write out the master equation for this 4-level system by treating $%
\widetilde{H}_{0}$ as unperturbed part, and $\widetilde{H}_{1}$ as
perturbation, the reduced master equation is\cite{DiVincenzo2005,Burkard2009}%
:
\begin{equation}
\frac{\partial \widetilde{\rho }(t)}{{\partial }t}=-i\left[ \widetilde{H}%
_{e},\widetilde{\rho }(t)\right] -\int_{0}^{t}d\,t^{\prime }\,X(t,t^{\prime
}).  \label{ME_S}
\end{equation}%
where $\widetilde{H}_{e}\equiv \sum_{i=0}^{3}\varepsilon _{i}|i\rangle
\langle i|$ and $X(t,t^{\prime })$ is
\begin{eqnarray*}
&&X(t,t^{\prime })\equiv \mathrm{Tr}_{bath}\left[ \widetilde{H}_{1},e^{-i%
\widetilde{H}_{0}t}\left[ \widetilde{H}_{1},\widetilde{\rho }(t{-t^{\prime }}%
)\right] e^{i\widetilde{H}_{0}t}\right] \\
&=&\sum_{k}g_{k}^{\prime 2}n_{k}e^{i\omega _{k}t}\left[ \varrho _{-}e^{-i%
\widetilde{H}_{e}t}\varrho _{+}\widetilde{\rho }(t{-t^{\prime }})e^{i%
\widetilde{H}_{e}t}-e^{-i\widetilde{H}_{e}t}\varrho _{+}\widetilde{\rho }(t{%
-t^{\prime }})e^{i\widetilde{H}_{e}t}\varrho _{-}\right] \\
&+&\sum_{k}g_{k}^{\prime 2}n_{k}e^{-i\omega _{k}t}\left[ e^{-i\widetilde{H}%
_{e}t}\widetilde{\rho }(t{-t^{\prime }})\varrho _{-}e^{i\widetilde{H}%
_{e}t}\varrho _{+}-\varrho _{+}e^{-i\widetilde{H}_{e}t}\widetilde{\rho }(t{%
-t^{\prime }})\varrho _{-}e^{i\widetilde{H}_{e}t}\right] \\
&+&\sum_{k}g_{k}^{\prime 2}(n_{k}+1)e^{-i\omega _{k}t}\left[ \varrho
_{+}e^{-i\widetilde{H}_{e}t}\varrho _{-}\widetilde{\rho }(t{-t^{\prime }}%
)e^{i\widetilde{H}_{e}t}-e^{-i\widetilde{H}_{e}t}\varrho _{-}\widetilde{\rho
}(t{-t^{\prime }})e^{i\widetilde{H}_{e}t}\varrho _{+}\right] \\
&+&\sum_{k}g_{k}^{\prime 2}(n_{k}+1)e^{i\omega _{k}t}\left[ e^{-i\widetilde{H%
}_{e}t}\widetilde{\rho }(t{-t^{\prime }})\varrho _{+}e^{i\widetilde{H}%
_{e}t}\varrho _{-}-\varrho _{-}e^{-i\widetilde{H}_{e}t}\widetilde{\rho }(t{%
-t^{\prime }})\varrho _{+}e^{i\widetilde{H}_{e}t}\right] .
\end{eqnarray*}

The above master equation is a $4\times 4$ matrix equation. According to the
Kronecker product property and technique to Lyapunov matrix equation in
matrix theory, by expanding the matrix $\rho (t)$ into vector $\overline{vec}%
[\rho (t)]$ along row, the equation becomes,
\begin{equation*}
\left\{ \frac{\partial }{{\partial }t}+i\left[ \widetilde{H}_{e}{\otimes I}%
_{4\times 4}-{I}_{4\times 4}{\otimes }\widetilde{H}_{e}\right] \right\}
\overline{vec}\left[ \widetilde{\rho }(t)\right] =-\sum_{k}g_{k}^{\prime
2}\int_{0}^{t}d\,t^{\prime }F_{k}(t{-t^{\prime }})\overline{vec}\left[
\widetilde{\rho }({t^{\prime }})\right] ,
\end{equation*}%
with%
\begin{eqnarray}
F_{k}(t) &\equiv &n_{k}e^{i\omega _{k}t}\left[ \left( \varrho _{-}e^{-i%
\widetilde{H}_{e}t}\varrho _{+}\right) {\otimes }\left( e^{i\widetilde{H}%
_{e}t}\right) ^{T}-\left( e^{-i\widetilde{H}_{e}t}\varrho _{+}\right) {%
\otimes }\left( e^{i\widetilde{H}_{e}t}\varrho _{-}\right) ^{T}\right]
\notag \\
&+&n_{k}e^{-i\omega _{k}t}\left[ e^{-i\widetilde{H}_{e}t}{\otimes }\left(
\varrho _{-}e^{i\widetilde{H}_{e}t}\varrho _{+}\right) ^{T}-\left( \varrho
_{+}e^{-i\widetilde{H}_{e}t}\right) {\otimes }\left( \varrho _{-}e^{i%
\widetilde{H}_{e}t}\right) ^{T}\right]  \notag \\
&+&(n_{k}+1)e^{-i\omega _{k}t}\left[ \left( \varrho _{+}e^{-i\widetilde{H}%
_{e}t}\varrho _{-}\right) {\otimes }\left( e^{i\widetilde{H}_{e}t}\right)
^{T}-\left( e^{-i\widetilde{H}_{e}t}\varrho _{-}\right) {\otimes }\left( e^{i%
\widetilde{H}_{e}t}\varrho _{+}\right) ^{T}\right]  \notag \\
&+&(n_{k}+1)e^{i\omega _{k}t}\left[ e^{-i\widetilde{H}_{e}t}{\otimes }\left(
\varrho _{+}e^{i\widetilde{H}_{e}t}\varrho _{-}\right) ^{T}-\left( \varrho
_{-}e^{-i\widetilde{H}_{e}t}\right) {\otimes }\left( \varrho _{+}e^{i%
\widetilde{H}_{e}t}\right) ^{T}\right] .  \label{FtDef}
\end{eqnarray}

In order to seek the non-Markovian effect of the non-linear bath, we solve
the above equation by using Laplace transformation, rather than by replacing
$\widetilde{\rho }({t^{\prime }})$ as $\widetilde{\rho }(t)$ which is the
usual treatment in the literature known as the Markovian approximation.
After the Laplace transformation and convolution theorem, the master
equation of the system can be obtained as:
\begin{equation}
U(P)_{16\times 16}\overline{vec}\left[ \overline{\widetilde{\rho }(P)}\right]
=\overline{vec}\left[ \widetilde{\rho }(0)\right]  \label{ME_Laplace}
\end{equation}%
with
\begin{equation*}
U(P)_{16\times 16}=PI_{16\times 16}+i\left[ \widetilde{H}_{e}{\otimes I}%
_{4\times 4}-{I}_{4\times 4}{\otimes }\widetilde{H}_{e}\right]
+\sum_{k}g_{k}^{\prime 2}\overline{F_{k}(P)}_{16\times 16},
\end{equation*}%
\ and $\overline{F_{k}(P)}_{16\times 16}$\ is the Laplace transformation of $%
F_{k}(t)_{16\times 16}$. The master equation (\ref{ME_Laplace}) leads to
following uncoupled equations:

\begin{eqnarray}
\left[ P+i{E_{p}}+\left( 2\,n_{k}+1\right) \left( \sin ^{2}\theta
B_{7+}+\sin ^{2}\theta B_{8+}\right) \right] \widetilde{\rho }_{14}(P) &=&%
\widetilde{\rho }_{14}(0)  \label{rho14} \\
\left[ P-i{E_{p}}+\left( 2\,n_{k}+1\right) \left( \sin ^{2}\theta
B_{7-}+\sin ^{2}\theta B_{8-}\right) \right] \widetilde{\rho }_{41}(P) &=&%
\widetilde{\rho }_{41}(0)  \label{rho41}
\end{eqnarray}%
\begin{equation}
A_{44}\cdot \left[
\begin{array}{cccc}
\widetilde{\rho }_{12}(P) & \widetilde{\rho }_{13}(P) & \widetilde{\rho }%
_{24}(P) & \widetilde{\rho }_{34}(P)%
\end{array}%
\right] ^{T}=\left[
\begin{array}{cccc}
\widetilde{\rho }_{12}(0) & \widetilde{\rho }_{13}(0) & \widetilde{\rho }%
_{24}(0) & \widetilde{\rho }_{34}(0)%
\end{array}%
\right] ^{T}  \label{ME44}
\end{equation}%
\begin{equation}
A_{44}^{\prime }\cdot \left[
\begin{array}{cccc}
\widetilde{\rho }_{21}(P) & \widetilde{\rho }_{31}(P) & \widetilde{\rho }%
_{42}(P) & \widetilde{\rho }_{43}(P)%
\end{array}%
\right] ^{T}=\left[
\begin{array}{cccc}
\widetilde{\rho }_{21}(0) & \widetilde{\rho }_{31}(0) & \widetilde{\rho }%
_{42}(0) & \widetilde{\rho }_{43}(0)%
\end{array}%
\right] ^{T}  \label{ME44p}
\end{equation}%
\begin{equation}
A_{66}\cdot \left[
\begin{array}{cccccc}
\widetilde{\rho }_{11}(P) & \widetilde{\rho }_{22}(P) & \widetilde{\rho }%
_{23}(P) & \widetilde{\rho }_{32}(P) & \widetilde{\rho }_{33}(P) &
\widetilde{\rho }_{44}(P)%
\end{array}%
\right] ^{T}=\left[
\begin{array}{cccccc}
\widetilde{\rho }_{11}(0) & \widetilde{\rho }_{22}(0) & \widetilde{\rho }%
_{23}(0) & \widetilde{\rho }_{32}(0) & \widetilde{\rho }_{33}(0) &
\widetilde{\rho }_{44}(0)%
\end{array}%
\right] ^{T}  \label{ME66}
\end{equation}%
with the explicit definition of $B_{7\pm }$, $B_{8\pm }$, $A_{44}$, $%
A_{44}^{\prime }$ and $A_{66}$ being defined in Appendix.

Before solving these equations, we first have a look at the initial
condition and the physical quantities. In this work, the physcial quantity
under our concern is the population difference $P(t)\equiv \langle \sigma
_{x}^{A}\rangle (t)=\mathrm{Tr}\left[ \rho (t)\sigma _{x}^{A}{\otimes I}%
_{2\times 2}\right] $. It can be rewritten as,%
\begin{equation}
P(t)\equiv \mathrm{Tr}\left[ \widetilde{\rho }(t)\widetilde{\sigma _{x}^{A}}%
\right]
\end{equation}%
where $\widetilde{\sigma _{x}^{A}}\equiv Te^{S}\sigma
_{x}^{A}e^{-S}T=T(\sigma _{x}^{A}\cos \theta _{0}+\sigma _{z}^{A}\sigma
_{x}^{B}\sin \theta _{0})T$, which is%
\begin{equation}
\widetilde{\sigma _{x}^{A}}=\left[
\begin{array}{cccc}
0 & \sin \left( \theta +{\theta _{0}}\right) & -\cos \left( \theta +{\theta
_{0}}\right) & 0 \\
\sin \left( \theta +{\theta _{0}}\right) & 0 & 0 & \cos \left( \theta +{%
\theta _{0}}\right) \\
-\cos \left( \theta +{\theta _{0}}\right) & 0 & 0 & \sin \left( \theta +{%
\theta _{0}}\right) \\
0 & \cos \left( \theta +{\theta _{0}}\right) & \sin \left( \theta +{\theta
_{0}}\right) & 0%
\end{array}%
\right]
\end{equation}%
thus,
\begin{eqnarray}
P(t) &=&\left( \widetilde{\rho }_{12}+\widetilde{\rho }_{34}+\widetilde{\rho
}_{21}+\widetilde{\rho }_{43}\right) \sin \left( \theta +\theta _{0}\right)
+\left( \widetilde{\rho }_{24}-\widetilde{\rho }_{13}+\widetilde{\rho }_{42}-%
\widetilde{\rho }_{31}\right) \cos \left( \theta +\theta _{0}\right)  \notag
\\
&=&2Re(\widetilde{\rho }_{12}+\widetilde{\rho }_{34})\sin \left( \theta
+\theta _{0}\right) +2Re(\widetilde{\rho }_{24}-\widetilde{\rho }_{13})\cos
\left( \theta +\theta _{0}\right) .  \label{P(t)WOIC}
\end{eqnarray}%
where the time dependant $\widetilde{\rho }(t)$ is replace by $\widetilde{%
\rho }$ for simplicity. From the above expression, we can see only Eq. (\ref%
{ME44}) is relevant to the dynamics of population difference. According to
Eq. (\ref{ME44}), we have
\begin{eqnarray}
\widetilde{\rho }_{12}(P)+\widetilde{\rho }_{34}(P) &=&{\frac{\,\left( P+{%
d_{3}}\right) \left[ {\widetilde{\rho }_{12}(0)}+{\widetilde{\rho }_{34}(0)}%
\right] +f\left[ {\widetilde{\rho }_{24}(0)}-{\widetilde{\rho }_{13}(0)}%
\right] }{\,\left( P+{d_{1}}\right) \left( P+{d_{3}}\right) -f^{2}}} \\
\widetilde{\rho }_{24}(P)-\widetilde{\rho }_{13}(P) &=&{\frac{\,\left( P+{%
d_{1}}\right) \left[ {\widetilde{\rho }_{24}(0)}-{\widetilde{\rho }_{13}(0)}%
\right] +f\left[ {\widetilde{\rho }_{12}(0)}+{\widetilde{\rho }_{34}(0)}%
\right] }{\,\left( P+{d_{1}}\right) \left( P+{d_{3}}\right) -f^{2}}}
\end{eqnarray}%
\begin{eqnarray}
d_{1} &=&i\left( {E_{p}}-{E_{m}}\right) /2+\sum_{k}\left( 2\,n_{k}+1\right)
\cos ^{2}\theta {B_{1+}} \\
d_{3} &=&i\left( \,{E_{p}}+\,{E_{m}}\right) /2+\sum_{k}\left(
2\,n_{k}+1\right) \sin ^{2}\theta {B_{1+}}
\end{eqnarray}%
\begin{equation}
f=\sum_{k}\left( 2\,n_{k}+1\right) \cos \left( \theta \right) \sin \left(
\theta \right) {B_{1+}}
\end{equation}%
Suppose the system is in the upper eigenstate of $\sigma _{x}^{A}$ and $%
\sigma _{x}^{B}$ at the initial time t=0, therefore, the initial condition
is given by
\begin{equation}
\rho (0)=\frac{1}{2}\left[
\begin{array}{cc}
1 & 1 \\
1 & 1%
\end{array}%
\right] \bigotimes \frac{1}{2}\left[
\begin{array}{cc}
1 & 1 \\
1 & 1%
\end{array}%
\right] =\frac{1}{4}\left[
\begin{array}{cccc}
1 & 1 & 1 & 1 \\
1 & 1 & 1 & 1 \\
1 & 1 & 1 & 1 \\
1 & 1 & 1 & 1%
\end{array}%
\right]
\end{equation}%
therefore, $\widetilde{\rho }(0)\equiv \frac{1}{4}Te^{S}(I_{2\times
2}^{A}+\sigma _{x}^{A})(I_{2\times 2}^{B}+\sigma _{x}^{B})e^{-S}T=\frac{1}{4}%
T(I_{2\times 2}^{A}+\sigma _{x}^{A}\cos \theta _{0}+\sigma _{z}^{A}\sigma
_{x}^{B}\sin \theta _{0})(I_{2\times 2}^{B}+\sigma _{x}^{B})T$, which is
\begin{equation*}
\widetilde{\rho }(0)=\frac{1}{4}\left[
\begin{array}{cccc}
1+\sin \left( {\theta _{0}}\right) & \cos \left( \theta \right) +\sin \left(
\theta +{\theta _{0}}\right) & \sin \left( \theta \right) -\cos \left(
\theta +{\theta _{0}}\right) & \cos \left( {\theta _{0}}\right) \\
\cos \left( \theta \right) +\sin \left( \theta +{\theta _{0}}\right) &
1+\sin \left( 2\,\theta +{\theta _{0}}\right) & -\cos \left( 2\,\theta +{%
\theta _{0}}\right) & \cos \left( \theta +{\theta _{0}}\right) +\sin \left(
\theta \right) \\
\sin \left( \theta \right) -\cos \left( \theta +{\theta _{0}}\right) & -\cos
\left( 2\,\theta +{\theta _{0}}\right) & 1-\sin \left( 2\,\theta +{\theta
_{0}}\right) & \sin \left( \theta +{\theta _{0}}\right) -\cos \left( \theta
\right) \\
\cos \left( {\theta _{0}}\right) & \cos \left( \theta +{\theta _{0}}\right)
+\sin \left( \theta \right) & \sin \left( \theta +{\theta _{0}}\right) -\cos
\left( \theta \right) & 1-\sin \left( {\theta _{0}}\right)%
\end{array}%
\right]
\end{equation*}%
from which we have
\begin{eqnarray}
{\widetilde{\rho }_{12}(0)}+{\widetilde{\rho }_{34}(0)} &=&\sin \left(
\theta +{\theta _{0}}\right) /2 \\
{\widetilde{\rho }_{24}(0)}-{\widetilde{\rho }_{13}(0)} &=&\cos \left(
\theta +{\theta _{0}}\right) /2
\end{eqnarray}%
Therefore, the population difference is
\begin{eqnarray}
P(t) &=&Re\left[ \mathscr{L}^{-1}\frac{\,\left( P+{d_{3}}\right) \sin
^{2}\Theta +f\sin \Theta \cos \Theta }{\,\left( P+{d_{1}}\right) \left( P+{%
d_{3}}\right) -f^{2}}\right]  \notag \\
&+&Re\left[ \mathscr{L}^{-1}{\frac{\,\left( P+{d_{1}}\right) \cos ^{2}\Theta
+f\sin \Theta \cos \Theta }{\,\left( P+{d_{1}}\right) \left( P+{d_{3}}%
\right) -f^{2}}}\right] .  \label{P(t)Laplace}
\end{eqnarray}%
where, $\mathscr{L}^{-1}$ is the Laplace inversion operator which corresponds to $\frac{1}{2\pi i}\int_{\sigma
-i\infty }^{\sigma +i\infty }dPe^{Pt}$, $\Theta \equiv \theta +{\theta _{0}}$
and
\begin{eqnarray}
d_{1} &=&i\left( {E_{p}}-{E_{m}}\right) /2+\cos ^{2}\theta G(P) \\
d_{3} &=&i\left( \,{E_{p}}+\,{E_{m}}\right) /2+\sin ^{2}\theta G(P)
\end{eqnarray}%
\begin{equation}
f=\cos \left( \theta \right) \sin \left( \theta \right) G(P)
\end{equation}%
\begin{equation}
G(P)=\int_{0}^{\infty }d\omega ^{\prime }\frac{J^{\prime }(\omega ^{\prime })%
}{{P+i\omega }^{\prime }}\coth (\beta \omega ^{\prime }),
\end{equation}%
with
\begin{equation}
J^{\prime }(\omega )=J(\omega )\left( \frac{\eta \Delta _{B}\cos
\theta _{0}}{\omega +\eta \Delta _{B}\cos \theta _{0}}\right) ^{2}
\end{equation}

Here we would like to summarize the
approximations we have made. Two approximations are made: The first one is
the omission of $H_{2}^{\prime }$ , which is a 4th order approximation to
the B-bath coupling. The second one is the Born approximation for deriving
the master equation (\ref{ME_S}), which is consistent with the first
approximation. Therefore, our treatment is applicable in low temperature for
$\alpha {\ll }1$ and $g_{0},T{\ll }\Delta _{A},\Delta _{B}$.

\section{Quasi-adiabatic path-integral method}

\label{seciv}

The QUAPI method is a numerical scheme based on a exact methodology \cite%
{Makarov1994,Makri1995a,Makri1995b,Thorwart2004}. The starting point of
QUAPI method is the generic system-bath Hamiltonian
\begin{equation*}
H=H_{0}+\sum_{j}\frac{P_{j}^{2}}{2m_{j}}+\frac{1}{2}m_{j}\omega
_{j}^{2}(Q_{j}-c_{j}s/m_{j}\omega _{j}^{2})^{2}.
\end{equation*}%
where, $H_{0}$ is the Hamiltonian for the bare system, $s$ is the system
coordinate, and $Q_{j}$ are harmonic bath coordinates which are linearly
coupled to the system coordinate. The characteristics of the bath are
captured in the spectral density function
\begin{equation}
J(\omega )=\frac{\pi }{2}\sum_{j}\frac{c_{j}^{2}}{m_{j}\omega _{j}}\delta
(\omega -\omega _{j}).
\end{equation}%
The reduced density matrix of the system evolve as $\rho (s^{\prime \prime
},s^{\prime },t)=\mathtt{Tr}_{bath}\left\langle s^{\prime \prime
}\right\vert e^{-iH_{0}t}\rho (0)e^{iH_{0}t}\left\vert s^{\prime
}\right\rangle $. If the path integral representation is discretized by $N$
time steps of length $\Delta t=t/N$ and the initial density matrix is
assumed to be $\rho (0)=\rho _{s}(0)\rho _{bath}(0)$, the reduced density
matrix takes the form%
\begin{eqnarray}
&&\rho (s^{\prime \prime },s^{\prime
},t)=\sum_{s_{N-1}^{+}}\sum_{s_{N-1}^{-}}\cdots
\sum_{s_{1}^{+}}\sum_{s_{1}^{-}}\sum_{s_{0}^{+}}\sum_{s_{0}^{-}}\left\langle
s^{\prime \prime }\right\vert e^{-iH_{0}\Delta t}\left\vert
s_{N-1}^{+}\right\rangle \cdots \left\langle s_{1}^{+}\right\vert
e^{-iH_{0}\Delta t}\left\vert s_{0}^{+}\right\rangle  \notag \\
&&\left\langle s_{0}^{+}\right\vert \rho _{s}(0)\left\vert
s_{0}^{-}\right\rangle \left\langle s_{0}^{-}\right\vert e^{iH_{0}\Delta
t}\left\vert s_{1}^{-}\right\rangle \cdots \left\langle
s_{N-1}^{-}\right\vert e^{iH_{0}\Delta t}\left\vert s^{\prime }\right\rangle
I(s_{0}^{+},s_{0}^{-},s_{1}^{+},s_{1}^{-},\cdots
,s_{N-1}^{+},s_{N-1}^{-},s^{\prime \prime },s^{\prime },\Delta t)  \notag \\
&&  \label{rhoss}
\end{eqnarray}%
where the discrete variable representation (DVR) is used, the symbol $%
s_{k}^{\pm }$ ($k=0.....N-1$) denotes the system coordinate at the time $%
k\Delta t$ on the forward and backward discretized Feynman path. $\left\vert
s_{k}^{\pm }\right\rangle $ ($k=0.....N-1$) are the eigenstates of the
system coordinate operator $s$. If a symmetric splitting of the
time-evolution operator is employed $e^{-iH\Delta t}=e^{-iH_{env}\Delta
t/2}e^{-iH_{0}\Delta t}e^{iH_{env}\Delta t/2}$ with $H_{env}=H-H_{0}$, the
corresponding influence functional reads%
\begin{eqnarray}
&&I(s_{0}^{+},s_{0}^{-},s_{1}^{+},s_{1}^{-},\cdots
,s_{N-1}^{+},s_{N-1}^{-},s^{\prime \prime },s^{\prime },\Delta t)  \notag \\
&=&\mathtt{Tr}_{bath}\left[ e^{-iH_{env}(s^{\prime \prime })\Delta
t/2}e^{-iH_{env}(s_{N-1}^{+})\Delta t}\cdots e^{-iH_{env}(s_{0}^{+})\Delta
t/2}\right.  \notag \\
&&\left. \times \rho _{bath}(0)e^{iH_{env}(s_{0}^{-})\Delta t/2}\cdots
e^{iH_{env}(s_{N-1}^{-})\Delta t}e^{iH_{env}(s^{\prime })\Delta t/2}\right] ,
\end{eqnarray}%
One can find that the equilibrium position of the bath mode is adiabatically
displaced along the system coordinate. If $H_{0}$ provides a reasonable
zeroth-order approximation to the dynamics, the quasi-adiabatic propagator
is accurate for fairly large time steps. That is the quasi-adiabatic
partitioning is a good representation when the bath property is mainly
adiabatic, where the bath can keep up with the motion the system quickly.
And the discrete path is to take into account of the non-adiabatic effect.
For most of case, the quasi-adiabatic partitioning is reasonable especially
when the system bath coupling is not strong. Therefore, the QUAPI
discretization permits fairly large time steps when the adiabatic bath
dominates the system dynamics. If the bath is purely adiabatic, even no
discretization is needed. In the continuous limit (that is for $\Delta {t}%
\rightarrow 0,N\rightarrow \infty $) the influence functional has been
calculated by Feynman and Vernon
\begin{eqnarray}
I &=&exp\left\{ -\frac{1}{\hbar }\int_{0}^{t}dt^{\prime }\int_{0}^{t^{\prime
}}dt^{\prime \prime }\left[ s^{+}(t^{\prime })-s^{-}(t^{\prime })\right]
\right.  \notag \\
&&\times \left[ \alpha (t^{\prime }-t^{\prime \prime })s^{+}(t^{\prime
\prime })-\alpha ^{\ast }(t^{\prime }-t^{\prime \prime })s^{-}(t^{\prime
\prime })\right]  \notag \\
&&\left. -\frac{i}{\hbar }\int_{0}^{t}dt^{\prime }\sum_{j}\frac{c_{j}^{2}}{%
2m_{j}\omega _{j}^{2}}\left[ s^{+}(t^{\prime })^{2}-s^{-}(t^{\prime })^{2}%
\right] \right\}  \label{I_FeynmanVernon}
\end{eqnarray}%
where $\alpha (t)$ is the bath response function, which can be expressed in
terms of the spectral density as
\begin{equation}
\alpha (t)=\frac{1}{\pi }\int_{0}^{\infty }{d\omega }J(\omega )\left[ \coth
\left( \frac{\beta \omega _{j}\hbar }{2}\cos (\omega _{j}t)-i\sin (\omega
_{j}t)\right) \right] .  \label{alphat}
\end{equation}%
The last term in Eq.(\ref{I_FeynmanVernon}) arises from the "counter-terms"
which are grouped with the bath Hamiltonian in the quasi-adiabatic splitting
of the propagator. With the quasi-adiabatic discretization of the path
integral, the influence functional, Eq. \ref{I_FeynmanVernon}, takes the form%
\begin{equation}
I=exp\left\{ -\frac{1}{\hbar }\sum\limits_{k=0}^{N}\sum\limits_{k^{\prime
}=0}^{k}\left[ s_{k}^{+}-s_{k}^{-}\right] \left[ \eta _{kk^{\prime
}}s_{k^{\prime }}^{+}-\eta _{kk^{\prime }}^{\ast }s_{k^{\prime }}^{-}\right]
\right\} ,  \notag
\end{equation}%
where $s_{N}^{+}=s^{\prime \prime }$ and $s_{N}^{-}=s^{\prime }$. The
coefficients $\eta _{kk^{\prime }}$ can be obtained by substituting the
discretized path into the Feynman-Vernon expression Eq.(\ref{I_FeynmanVernon}%
), which is given in Ref. \cite{Makri1995a}.

The QUAPI method is essentially a tensor multiplication scheme, which
exploits the observation that for environments characterized by broad
spectra the response function $\alpha (t)$ decays within a finite time
interval. From the expression of the Feynman and Vernon influence funcitonal
Eq. (\ref{rhoss}), one can see that $\alpha (t)$ characterizes nonlocal
interactions, which connects system coordinate $s(t^{\prime })$ with $%
s(t^{\prime \prime })$. The path $s^{\pm }(t^{\prime })$ at time $t^{\prime
} $ is connected to the all the paths $s^{-}(t^{\prime \prime })$ at earlier
times, which makes the evaluation of Eq. (\ref{rhoss}) a hard task. However,
for a bath with a broad spectral density, such as a power law distribution
of the spectral density, $\alpha (t)$ has the finite memory, the memory
length typically extending over only a few time slices when the
quasi-adiabatic propagator is used to discretize the path integral. After
discarding the negligible "long-distance interaction" with $t^{\prime
}-t^{\prime \prime }>\Delta k_{\max }\Delta t$ (or $k-k^{\prime }>\Delta
k_{\max }$), the resulting path integral can be evaluated iteratively by
multiplication of a tensor of rank $2\Delta k_{max}$. In other words, there
exists an augmented reduced density tensor of rank $\Delta k_{max}$ that
obeys Markovian dynamics. The details of the multiplication scheme is
discussed to a great extent in the literature, here we only present the
essential parameters and mention briefly how to adopt it to our specific
problem \cite{Makarov1994,Makri1995a,Makri1995b,Thorwart2004}.Here we discuss briefly the parameters used in the
QUAPI method:

(i)The first parameter time-step $\Delta {t}$ used for the quasi-adiabatic
splitting of the path-integral. The memory time of the non-Markovian steps
used by QUAPI is $\Delta {k_{max}}\Delta {t}$. The stability of the
iterative density matrix propagation ensures the choices of $\Delta {t}$, it
should not be too big nor too small, since the non-adiabatic effect requires
more splitting of the path integral, that is smaller $\Delta {t}$. Whereas,
since the memory length $\Delta {k_{max}}\Delta {t}$ is usually a fixed
value for a particular bath, QUAPI method prefers larger $\Delta {t}$, and
consequently smaller $\Delta {k_{max}}$ in consideration of the
numerical efficiency (note that the algorithm scales exponentially with $\Delta
k_{\max }$, also see the discussion of the second parameter $\Delta k_{\max }$).
Therefore, we should choose appropriate $\Delta {t}$ to take into account
both the non-adiabatic effect which prefer smaller time splitting and the
non-Markov effect which prefer long memory time, typically, we choose $%
\Delta t$ around $\frac{2\pi }{20\Delta_A }$, that is to choose tens of
fraction of the cycle time of the bare system dynamics.

(ii) The second parameter is the memory steps $\Delta {k_{max}}$. If $\Delta {k_{max}}\leq 1$, the dynamics is purely Markovian. If the non-locality extends over longer time,
terms with $\Delta {k_{max}}>1$ have to be included to obtain accurate
results. In order to acquire converge result, in the practical
implementation of QUAPI, one usually need to choose $\Delta k_{\max }$ large
enough so that the response function reduces to negligible value within the
length of $\Delta {k_{max}}\Delta {t}$. However this is a hard task, Since
augmented propagator tensor $A^{(\Delta {k_{max}})}$ is a vector of
dimension $(M^{2})^{\Delta {k_{max}}}$ ($M$ is the system dimension which is
four here), and the corresponding tensor propagator $T^{(2\Delta {k_{max}})}$
is a matrix of dimension $(M^{2})^{2\Delta {k_{max}}}$, the QUAPI scheme
scales exponentially with the parameter $\Delta k_{\max }$. Thus one can not
proceed the QUAPI calculation with very large $\Delta k_{\max }$, and
usually $\Delta k_{\max }$ is chosen less than 5 for $M=4$, and even
smaller for larger $M$.

\section{Results and discussion}

We report $P(t)$ as a
function of time in Fig.~1 to Fig.~4 for $\Delta _{A}=\Delta _{B}=0.1\omega
_{l}$, $g_{0}=0.1\Delta _{A}$ and $\omega _{d}=0.05\omega _{l}$. The analytical results are depicted in solid lines,
and QUAPI results of different $\Delta k$'s are scatters with different colors and shapes. In Fig.~1, where $%
\alpha $ is larger than $g_{0}/\Delta _{A}$ which is set to be $\alpha =0.3$%
, it shows that the decoherence is reduced by increasing the bath
temperature $T$ as predicted in Ref.~\cite{Montina2008}. However, in Fig.~2,
where $\alpha =0.01$, the coherence is not meliorated but rather damaged
with increasing $T$. Similarly, in Fig.~3, decoherence reduction with bath coupling when $\alpha =0.3$ is possible, whereas, in Fig.~4, there is only decoherence enhancing where $\alpha =0.01$.

One can see both analytical and numerical shows good agreement. When the coupling is small, e.g. $\alpha=0.01, 0.02$ and $0.03$ in Fig.~2 and Fig.~4, the two methods show perfect agreement, all the QUAPI results converges to our TRWA results as $\Delta k$ increases. When $\alpha$ becomes larger, e.g. $\alpha=0.3, 0.4$ and $0.5$, discrepancy appears, we attribute this discrepancy to that the QUAPI is not converged completely. Since when $\alpha$ becomes large the non-adiabatic boson contributes significantly, one need small $\Delta t$ to take into account of the non-adiabatic bosons. Whereas, the memory length is almost the same, consequently, one can not converge within $\Delta k=3$ which is the upper limit of our computation resources (note the algorithm scales exponentially as $\Delta k$). Finally, as a check of our analytical method, we report the strong qubit-TL coupling case, where $g_0=\Delta$. One can see perfect agreement is achieved between the analytical and the numerical methods.

\section{Conclusion}

In conclusion, without making RWA and Markov approximation, the dynamics of
a qubit coupled with a spin-boson bath is investigated. We proposed an unitary transformation to treat this problem. The results of our analytical method show good agreement with QUAPI, even when the qubit-TL coupling is as strong as the TL spacing. And checked the decoherence behavior with varying bath temperature $T$ and TLF-bath coupling.
The decoherence of TSS-A can be reduced increasing bath temperature $T$ or with increasing TL-bath coupling $\alpha$ only when the A-B coupling is smaller than B-bath coupling.

\section{acknowledgement}

Part of the work of P. Huang is done when he visit IMSS, KEK in Japan. We gratefully
acknowledge the support by China Scholarship Council, by the National Natural Science Foundation of China
(Grant No.10734020) and the National Basic Research Program of China (Grant No. 2011CB922202) and by
the Next Generation Supercomputer Project, Nanoscience Program, MEXT, Japan.

\appendix

\section{The matrix form and the matrix elements of $\overline{%
F_{k}(P)}_{16\times 16}$ and $U(P)_{16\times 16}$}

From Eq.~(\ref{FtDef}), the laplace transform of $F_{k}(t)$ is of following
form,
\begin{equation}
\left[
\begin{array}{cccccccccccccccc}
F_{{1,1}} & 0 & 0 & 0 & 0 & {F}_{{1,6}} & F_{{1,7}} & 0 & 0 & F_{{1,10}} &
F_{{1,11}} & 0 & 0 & 0 & 0 & 0 \\
0 & F_{{2,2}} & F_{{2,3}} & 0 & 0 & 0 & 0 & F_{{2,8}} & 0 & 0 & 0 & F_{{2,12}%
} & 0 & 0 & 0 & 0 \\
0 & F_{{3,2}} & F_{{3,3}} & 0 & 0 & 0 & 0 & F_{{3,8}} & 0 & 0 & 0 & F_{{3,12}%
} & 0 & 0 & 0 & 0 \\
0 & 0 & 0 & {F}_{{4,4}} & 0 & 0 & 0 & 0 & 0 & 0 & 0 & 0 & 0 & 0 & 0 & 0 \\
0 & 0 & 0 & 0 & {F}_{{5,5}} & 0 & 0 & 0 & F_{{5,9}} & 0 & 0 & 0 & 0 & F_{{%
5,14}} & F_{{5,15}} & 0 \\
F_{{6,1}} & 0 & 0 & 0 & 0 & F_{{6,6}} & F_{{6,7}} & 0 & 0 & F_{{6,10}} & 0 &
0 & 0 & 0 & 0 & F_{{6,16}} \\
F_{{7,1}} & 0 & 0 & 0 & 0 & F_{{7,6}} & F_{{7,7}} & 0 & 0 & 0 & F_{{7,11}} &
0 & 0 & 0 & 0 & F_{{7,16}} \\
0 & F_{{8,2}} & F_{{8,3}} & 0 & 0 & 0 & 0 & F_{{8,8}} & 0 & 0 & 0 & F_{{8,12}%
} & 0 & 0 & 0 & 0 \\
0 & 0 & 0 & 0 & F_{{9,5}} & 0 & 0 & 0 & F_{{9,9}} & 0 & 0 & 0 & 0 & F_{{9,14}%
} & {F}_{{9,15}} & 0 \\
F_{{10,1}} & 0 & 0 & 0 & 0 & F_{{10,6}} & 0 & 0 & 0 & F_{{10,10}} & F_{{10,11%
}} & 0 & 0 & 0 & 0 & F_{{10,16}} \\
F_{{11,1}} & 0 & 0 & 0 & 0 & 0 & F_{{11,7}} & 0 & 0 & F_{{11,10}} & F_{{11,11%
}} & 0 & 0 & 0 & 0 & F_{{11,16}} \\
0 & F_{{12,2}} & F_{{12,3}} & 0 & 0 & 0 & 0 & F_{{12,8}} & 0 & 0 & 0 & F_{{%
12,12}} & 0 & 0 & 0 & 0 \\
0 & 0 & 0 & 0 & 0 & 0 & 0 & 0 & 0 & 0 & 0 & 0 & F_{{13,13}} & 0 & 0 & 0 \\
0 & 0 & 0 & 0 & F_{{14,5}} & 0 & 0 & 0 & F_{{14,9}} & 0 & 0 & 0 & 0 & F_{{%
14,14}} & F_{{14,15}} & 0 \\
0 & 0 & 0 & 0 & F_{{15,5}} & 0 & 0 & 0 & F_{{15,9}} & 0 & 0 & 0 & 0 & F_{{%
15,14}} & F_{{15,15}} & 0 \\
0 & 0 & 0 & 0 & 0 & F_{{16,6}} & F_{{16,7}} & 0 & 0 & F_{{16,10}} & F_{{16,11%
}} & 0 & 0 & 0 & 0 & {F}_{{16,16}}%
\end{array}%
\right]
\end{equation}%
where $F_{k}(P)_{i,j}$ are expressed as $F_{i,j}$ for simplicity. In order
to save space, we do not give the explicit expressions of $F_{i,j}$ here. On
the other hand, $\left[ \widetilde{H}_{e}{\otimes I}_{4\times 4}-{I}%
_{4\times 4}{\otimes }\widetilde{H}_{e}\right] $ is just a diagonal matrix.
Therefore, from the definition of $U(P)_{16\times 16}$,
\begin{equation}
U(P)_{16\times 16}=PI_{16\times 16}+i\left[ \widetilde{H}_{e}{\otimes I}%
_{4\times 4}-{I}_{4\times 4}{\otimes }\widetilde{H}_{e}\right]
+\sum_{k}g_{k}^{\prime 2}\overline{F_{k}(P)}_{16\times 16},
\end{equation}%
we can get each element of $U(P)_{16\times 16}$ as,
\begin{equation*}
\left\{
\begin{array}{ccl}
U_{m,n} & = & \sum_{k}g_{k}^{\prime 2}F_{k}(P)_{m,n}\text{\ \ \ \ \ \ \ \ \
\ \ \ \ \ \ \ \ \ \ when }m\neq n \\
U_{m,m} & = & P+i\left[ \widetilde{H}_{e}{\otimes I}_{4\times 4}-{I}%
_{4\times 4}{\otimes }\widetilde{H}_{e}\right] _{m,m}+\sum_{k}g_{k}^{\prime
2}F_{k}(P)_{m,m}%
\end{array}%
\right.
\end{equation*}%
With the particular form of $U(P)_{16\times 16}$ one can decouple the master
equation Eq.~(\ref{ME_Laplace}) into equation sets with smaller dimension as shown
in Eq.~(\ref{rho14})-(\ref{ME66}), which are explicitly expressed in the
subsequenct Appendix. The parameters $B_{\pm }$'s which will be used in the
master equation sets are defined as,
\begin{eqnarray*}
B_{1\pm } &=&\frac{1}{{P\pm i\omega }_{k}},B_{2\pm }=\frac{1}{{P\pm i({%
\omega }_{k}-E}_{m}{)}}, \\
B_{3\pm } &=&\frac{1}{{P\pm i({\omega }_{k}+E}_{m}{)}},B_{4\pm }=\frac{1}{{%
P\pm i({\omega }_{k}-E}_{p}{)}}, \\
B_{5\pm } &=&\frac{1}{{P\pm i({\omega }_{k}-}\frac{{E}_{p}+E_{m}}{2}{)}}%
,B_{6\pm }=\frac{1}{{P\pm i({\omega }_{k}-}\frac{{E}_{p}-E_{m}}{2}{)}}, \\
B_{7\pm } &=&\frac{1}{{P\pm i({\omega }_{k}+}\frac{{E}_{p}-E_{m}}{2}{)}}%
,B_{8\pm }=\frac{1}{{P\pm i({\omega }_{k}+}\frac{{E}_{p}+E_{m}}{2}{)}}
\end{eqnarray*}

\section{Solve the $4\times4$ Master equation}

The master equation for $\rho _{12}$, $\rho _{13}$, $\rho _{24}$ and $\rho
_{34}$ is
\begin{equation}
A_{44}\cdot \left[
\begin{array}{cccc}
\rho _{12}(P) & \rho _{13}(P) & \rho _{24}(P) & \rho _{34}(P)%
\end{array}%
\right] ^{T}=\left[
\begin{array}{cccc}
\rho _{12}(0) & \rho _{13}(0) & \rho _{24}(0) & \rho _{34}(0)%
\end{array}%
\right] ^{T}
\end{equation}%
where\ $A_{44}$ is defined as%
\begin{equation}
\left[
\begin{array}{cccc}
P+{d_{1}}+n_{k}^{(1)}{d_{2k}} & -\sum_{k}\frac{{e_{1k}+}\,{e_{2k}}}{2} &
-n_{k}{e_{1k}} & -n_{k}{d_{2k}} \\
-\sum_{k}\frac{\,{e_{1k}}+\,{e_{2k}}}{2} & P+{d_{3}}+n_{k}^{(1)}{d_{4k}} &
n_{k}{d_{4k}} & n_{k}{e_{1k}} \\
-n_{k}^{(1)}{e_{1k}} & n_{k}^{(1)}{d_{4k}} & P+{d_{3}}+n_{k}{d_{4k}} &
-\sum_{k}\frac{\,{e_{1k}}-\,{e_{2k}}}{2} \\
-n_{k}^{(1)}{d_{2k}} & n_{k}^{(1)}{e_{1k}} & -\sum_{k}\frac{\,{e_{1k}}-\,{%
e_{2k}}}{2} & P+{d_{1}}+n_{k}{d_{2k}}%
\end{array}%
\right]  \label{A44}
\end{equation}%
where double indexes $k$ indicate a sum over $k$, $n_{k}^{(1)}\equiv n_{k}+1$
and
\begin{eqnarray}
d_{1} &=&i\left( {E_{p}}-{E_{m}}\right) /2+\sum_{k}g_{k}^{\prime 2}\left(
2\,n_{k}+1\right) \cos ^{2}\theta {B_{1+}} \\
d_{3} &=&i\left( \,{E_{p}}+\,{E_{m}}\right) /2+\sum_{k}g_{k}^{\prime
2}\left( 2\,n_{k}+1\right) \sin ^{2}\theta {B_{1+}}
\end{eqnarray}%
\begin{eqnarray}
d_{2k} &=&g_{k}^{\prime 2}\sin ^{2}\theta \left( {B_{2+}}+{B_{4-}}\right) \\
d_{4k} &=&g_{k}^{\prime 2}\cos ^{2}\theta \left( {B_{3+}}+{B_{4-}}\right) \\
e_{1k} &=&g_{k}^{\prime 2}\cos \left( \theta \right) \sin \left( \theta
\right) \left( {B_{4-}}+{B_{1+}}\right) \\
e_{2k} &=&g_{k}^{\prime 2}\left( 2\,n_{k}+1\right) \cos \left( \theta
\right) \sin \left( \theta \right) \left( {B_{4-}}-{B_{1+}}\right)
\end{eqnarray}%
By solving the above $4\times4$ linear algebra equation, we can get
\begin{eqnarray}
\widetilde{\rho }_{12}(P)+\widetilde{\rho }_{34}(P) &=&{\frac{\,\left( P+{%
d_{3}}\right) \left[ {\widetilde{\rho }_{12}(0)}+{\widetilde{\rho }_{34}(0)}%
\right] +f\left[ {\widetilde{\rho }_{24}(0)}-{\widetilde{\rho }_{13}(0)}%
\right] }{\,\left( P+{d_{1}}\right) \left( P+{d_{3}}\right) -f^{2}}} \\
\widetilde{\rho }_{24}(P)-\widetilde{\rho }_{13}(P) &=&{\frac{\,\left( P+{%
d_{1}}\right) \left[ {\widetilde{\rho }_{24}(0)}-{\widetilde{\rho }_{13}(0)}%
\right] +f\left[ {\widetilde{\rho }_{12}(0)}+{\widetilde{\rho }_{34}(0)}%
\right] }{\,\left( P+{d_{1}}\right) \left( P+{d_{3}}\right) -f^{2}}}
\end{eqnarray}%
\begin{equation*}
f\equiv \sum_{k}\left[ \left( 2\,n_{k}+1\right) {e_{1k}}-{e_{2k}}\right]
/2=\sum_{k}g_{k}^{\prime 2}\left( 2\,n_{k}+1\right) \cos \left( \theta
\right) \sin \left( \theta \right) {B_{1+}}
\end{equation*}%
As for the other $4\times 4$ master equation, since $\rho _{21}$, $\rho
_{31} $, $\rho _{42}$ and $\rho _{43}$ is just the complex conjugate of $%
\rho _{12} $, $\rho _{13}$, $\rho _{24}$ and $\rho _{34}$, we do not need to
solve them, and ,on the other hand, $A_{44}^{\prime }$ is the very similar
to $A_{44}$, the only difference is that $B_{+}$'s are changed to $B_{-}$'s
and vice versa.

\section{Solve the $6\times6$ Master equation}

The master equation for $\rho _{11}$, $\rho _{22}$, $\rho _{23}$, $\rho
_{32} $, $\rho _{33}$ and $\rho _{44}$ is
\begin{equation}
A_{66}\cdot \left[
\begin{array}{cccccc}
\widetilde{\rho }_{11}(P) & \widetilde{\rho }_{22}(P) & \widetilde{\rho }%
_{23}(P) & \widetilde{\rho }_{32}(P) & \widetilde{\rho }_{33}(P) &
\widetilde{\rho }_{44}(P)%
\end{array}%
\right] ^{T}=\left[
\begin{array}{cccccc}
\widetilde{\rho }_{11}(0) & \widetilde{\rho }_{22}(0) & \widetilde{\rho }%
_{23}(0) & \widetilde{\rho }_{32}(0) & \widetilde{\rho }_{33}(0) &
\widetilde{\rho }_{44}(0)%
\end{array}%
\right] ^{T}
\end{equation}%
\ $A_{66}$ defined as%
\begin{equation}
\left[
\begin{array}{cccccc}
P+n_{k}^{(1)}\left( a_{1k}+a_{2k}\right) & -n_{k}a_{1k} & -n_{k}b_{1k} &
-n_{k}b_{2k} & -n_{k}a_{2k} & 0 \\
-n_{k}^{(1)}a_{1k} & P+n_{k}a_{1k}+n_{k}^{(1)}a_{2k} & -\frac{b_{1}+b_{3}}{2}
& -\,\frac{b_{2}+\,b_{4}}{2} & 0 & -n_{k}a_{2k} \\
-n_{k}^{(1)}b_{1k} & -\frac{b_{1}+b_{3}}{2} & P+a_{3} & 0 & -\frac{%
b_{1}-b_{3}}{2} & n_{k}b_{1k} \\
-n_{k}^{(1)}b_{2k} & -\frac{b_{2}+\,b_{4}}{2} & 0 & P+a_{4} & -\frac{%
b_{2}-\,b_{4}}{2} & n_{k}b_{2k} \\
-n_{k}^{(1)}a_{2k} & 0 & -\frac{b_{1}-b_{3}}{2} & -\frac{b_{2}-\,b_{4}}{2} &
P+n_{k}^{(1)}a_{1k}+n_{k}a_{2k} & -n_{k}a_{1k} \\
0 & -n_{k}^{(1)}a_{2k} & n_{k}^{(1)}b_{1k} & n_{k}^{(1)}b_{2k} &
-n_{k}^{(1)}a_{1k} & P+n_{k}\left( a_{1k}+a_{2k}\right)%
\end{array}%
\right]  \label{A66}
\end{equation}%
where $b_{1}=\sum_{k}b_{1k}$, $b_{2}=\sum_{k}b_{2k}$, $b_{3}=\sum_{k}b_{3k}$%
, $b_{4}=\sum_{k}b_{4k}$,
\begin{eqnarray}
a_{1k} &=&g_{k}^{\prime 2}\cos ^{2}\theta \left( {B_{6+}}+{B_{6-}}\right) \\
a_{2k} &=&g_{k}^{\prime 2}\sin ^{2}\theta \left( {B_{5+}}+{B_{5-}}\right) \\
a_{3} &=&i{E_{m}}+\sum_{k}g_{k}^{\prime 2}\left( 2n_{k}+1\right) \left( \cos
^{2}\theta {B_{5-}}+\sin ^{2}\theta {B_{6+}}\right) \\
a_{4} &=&-i{E_{m}}+\sum_{k}g_{k}^{\prime 2}\left( 2n_{k}+1\right) \left(
\sin ^{2}\theta {B_{6-}}+\cos ^{2}\theta {B_{5+}}\right) \\
b_{1k} &=&g_{k}^{\prime 2}\cos \left( \theta \right) \sin \left( \theta
\right) \left( {B_{5-}}+{B_{6+}}\right) \\
b_{2k} &=&g_{k}^{\prime 2}\cos \left( \theta \right) \sin \left( \theta
\right) \left( {B_{6-}}+{B_{5+}}\right) \\
b_{3k} &=&g_{k}^{\prime 2}\left( 2n_{k}+1\right) \cos \left( \theta \right)
\sin \left( \theta \right) \left( {B_{5-}}-{B_{6+}}\right) \\
b_{4k} &=&g_{k}^{\prime 2}\left( 2n_{k}+1\right) \cos \left( \theta \right)
\sin \left( \theta \right) \left( {B_{6-}}-{B_{5+}}\right)
\end{eqnarray}%
The analytical solution of this equation set is still out of our capability, however, it is irrelevant to the quantities
which we are interested in. If we add the first two and last two lines of the master equation, we can find
\begin{equation}
\widetilde{\rho }_{11}(P)+\widetilde{\rho }_{22}(P)+\widetilde{\rho }%
_{33}(P)+\widetilde{\rho }_{44}(P)=1/P
\end{equation}%
which ensures the conservation of the trace of the reduced density operator $Tr\rho =1$.

\bibliographystyle{apsrev}
\bibliography{article}

%
%

\section*{Figures Captions}

{Fig.~1:} {$P(t)$ is plotted as a function of time for different temperatures $T/\Delta_A=1, 5, 10$ with analytical method (solid lines) and the QUAPI methods with $\Delta k=1, 2$ and $3$ (respectively black square, red circle and green triangle). The decoherence is reduced with temperature $T$. The parameters: $\Delta_A =\Delta_B $, $g_{0}=0.2\Delta_A $, $\alpha=0.3$, and the QUAPI parameter $\Delta t=0.4/\Delta_A$.}\newline

{Fig.~2:} {$P(t)$ is plotted as a function of time different temperatures $T/\Delta_A=0.1, 1, 5$ with analytical method (solid lines) and the QUAPI methods with $\Delta k=1, 2$ and $3$ (respectively black square, red circle and green triangle). The decoherence is enhanced with $T$. Two frequencies are dominating the dynamics which agree with the results of Jaynes-Cummings model. The parameters: $\Delta_A =\Delta_B $, $g_{0}=0.2\Delta_A $, $\alpha=0.01$, and the QUAPI parameter $\Delta t=0.4/\Delta_A$.}\newline

{Fig.~3:} {$P(t)$ is plotted as a function of time for different TL-bath coupling $\alpha=0.3, 0.4$ and $0.5$ with analytical method (solid lines) and the QUAPI methods with $\Delta k=1, 2$ and $3$ (respectively black square, red circle and green triangle). The decoherence is reduced with temperature $T$. The parameters: $\Delta_A =\Delta_B $, $g_{0}=0.2\Delta_A $, $T=0.1\Delta_A$, and the QUAPI parameter $\Delta t=0.4/\Delta_A$.}

{Fig.~4:} {$P(t)$ is plotted as a function of time for different TL-bath coupling $\alpha=0.01, 0.02$ and $0.03$ with analytical method (solid lines) and the QUAPI methods with $\Delta k=1, 2$ and $3$ (respectively black square, red circle and green triangle). The decoherence is reduced with temperature $T$. The parameters: $\Delta_A =\Delta_B $, $g_{0}=0.2\Delta_A $, $T=0.1\Delta_A$, and the QUAPI parameter $\Delta t=0.4/\Delta_A$.}

{Fig.~5:} {$P(t)$ is plotted as a function of time for strong qubit-TL coupling case $g_0=\Delta_A$ with analytical method (solid lines) and the QUAPI methods with $\Delta k=1, 2$ and $3$ (respectively black square, red circle and green triangle). The parameters: $\Delta_A =\Delta_B $, $\alpha=0.01$, and the QUAPI parameter $\Delta t=0.4/\Delta_A$.}

\clearpage
\begin{figure}[tbp]
\includegraphics[scale=0.8]{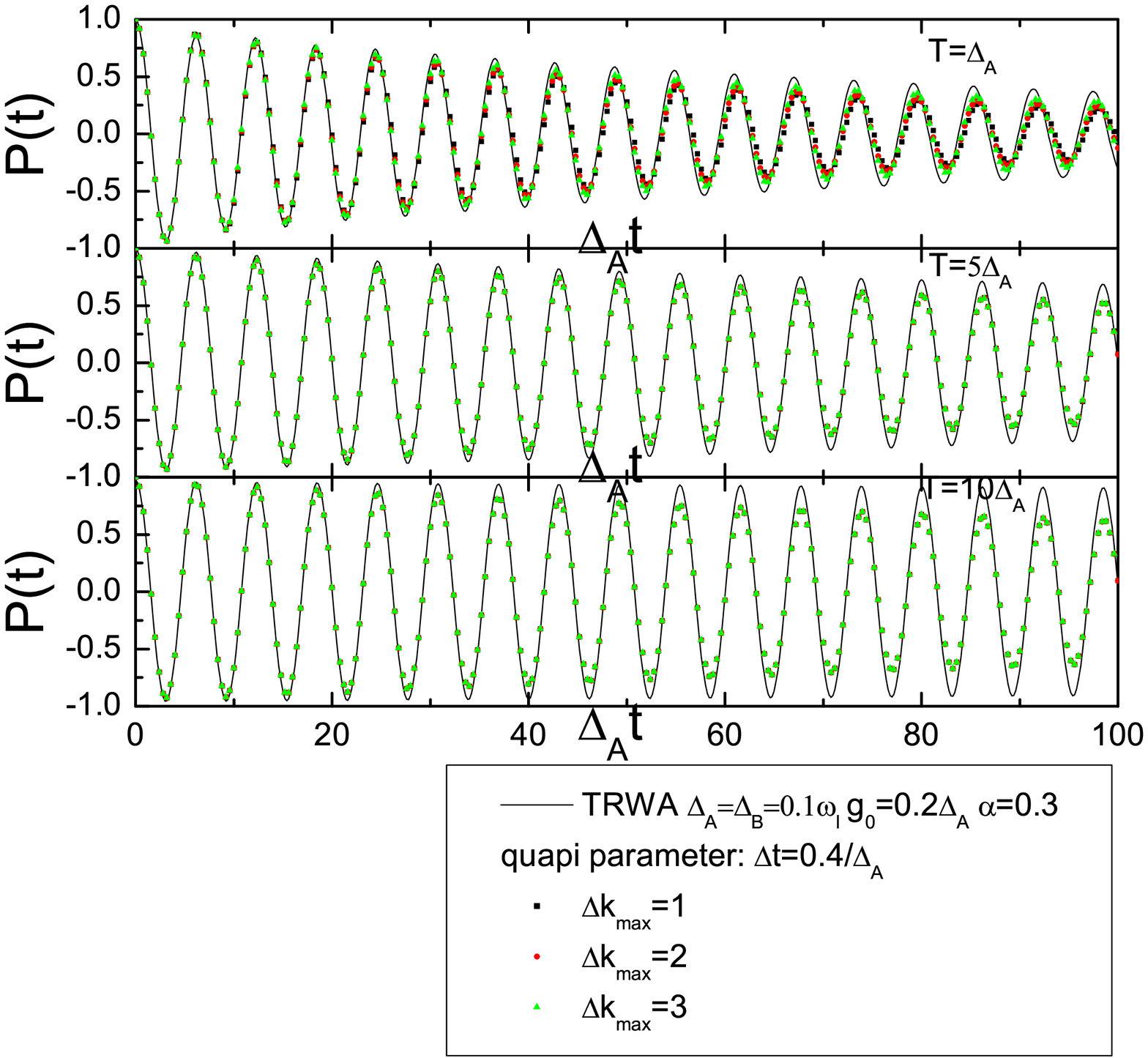}
\end{figure}
\clearpage
\begin{figure}[tbp]
\includegraphics[scale=0.8]{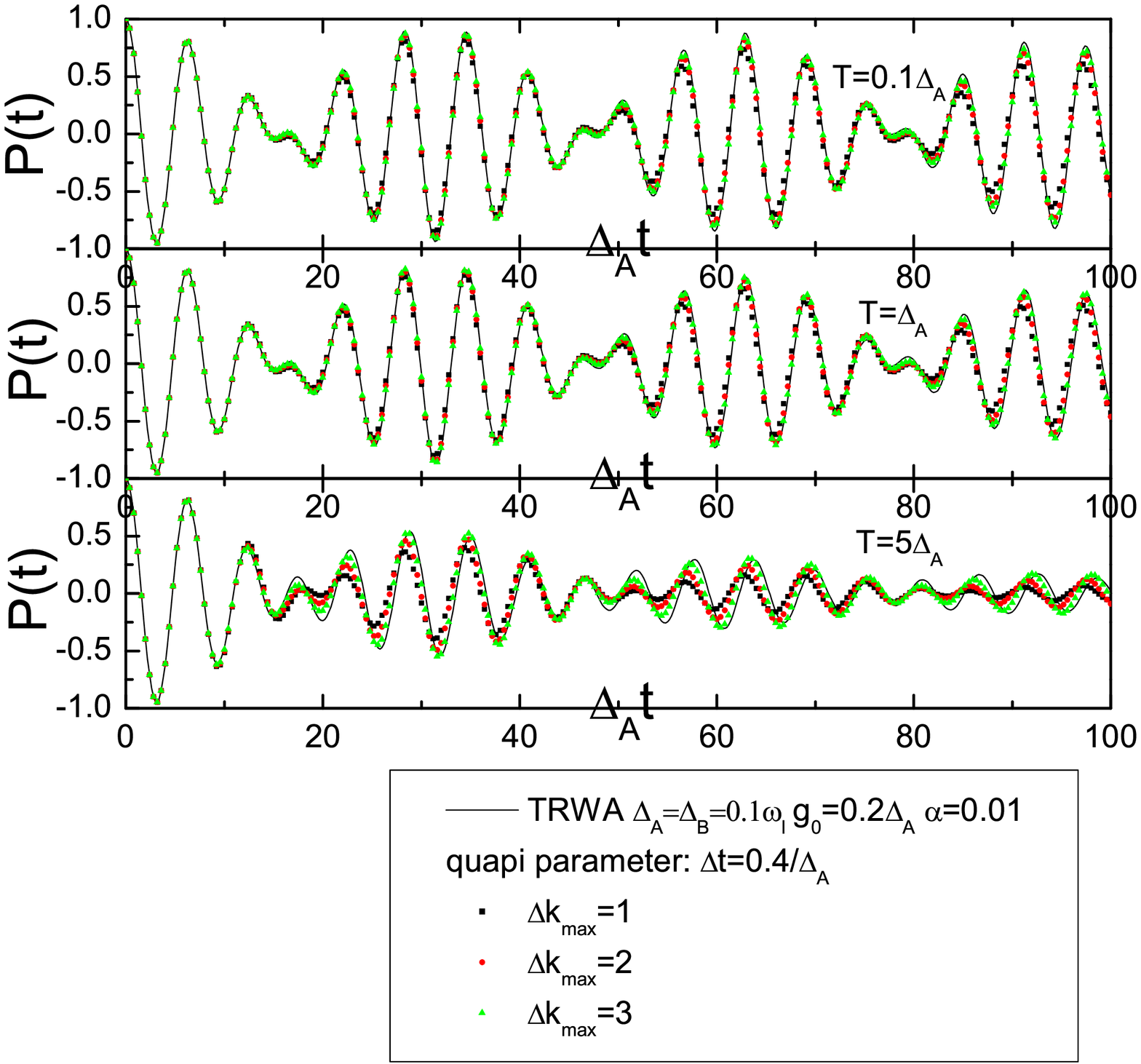}
\end{figure}
\clearpage
\begin{figure}[tbp]
\includegraphics[scale=0.8]{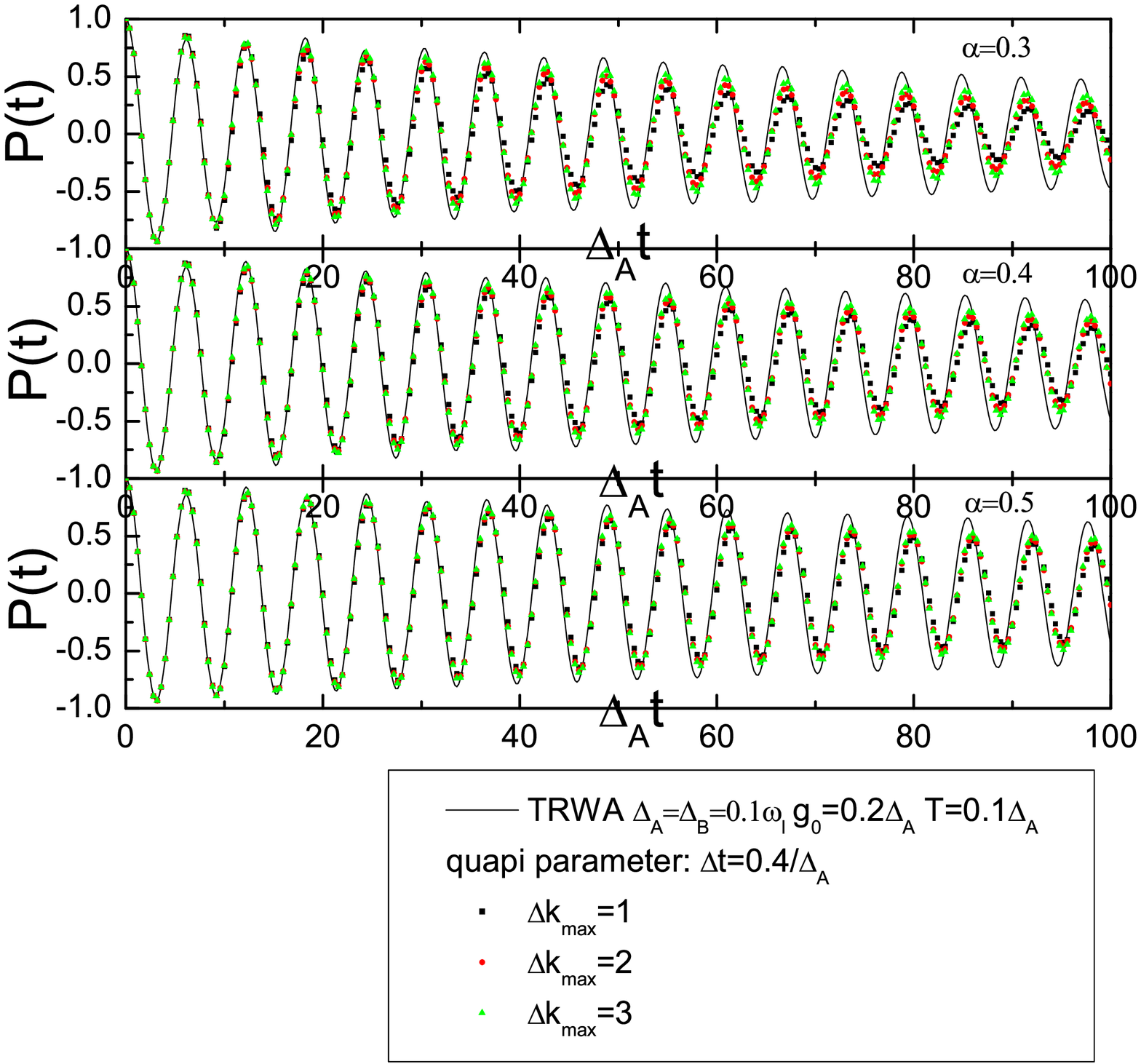}
\end{figure}
\clearpage
\begin{figure}[tbp]
\includegraphics[scale=0.8]{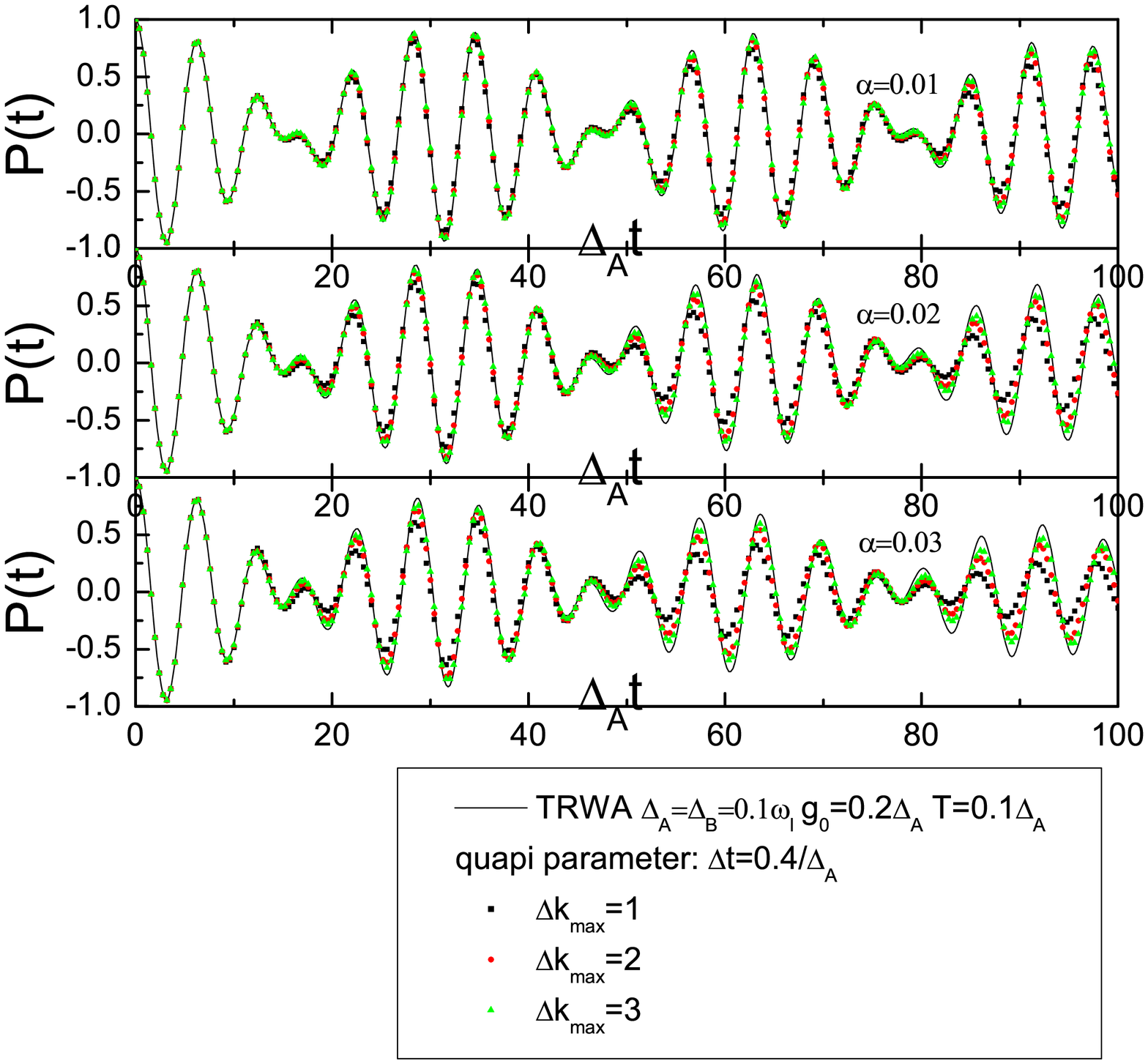}
\end{figure}
\clearpage
\begin{figure}[tbp]
\includegraphics[scale=0.8]{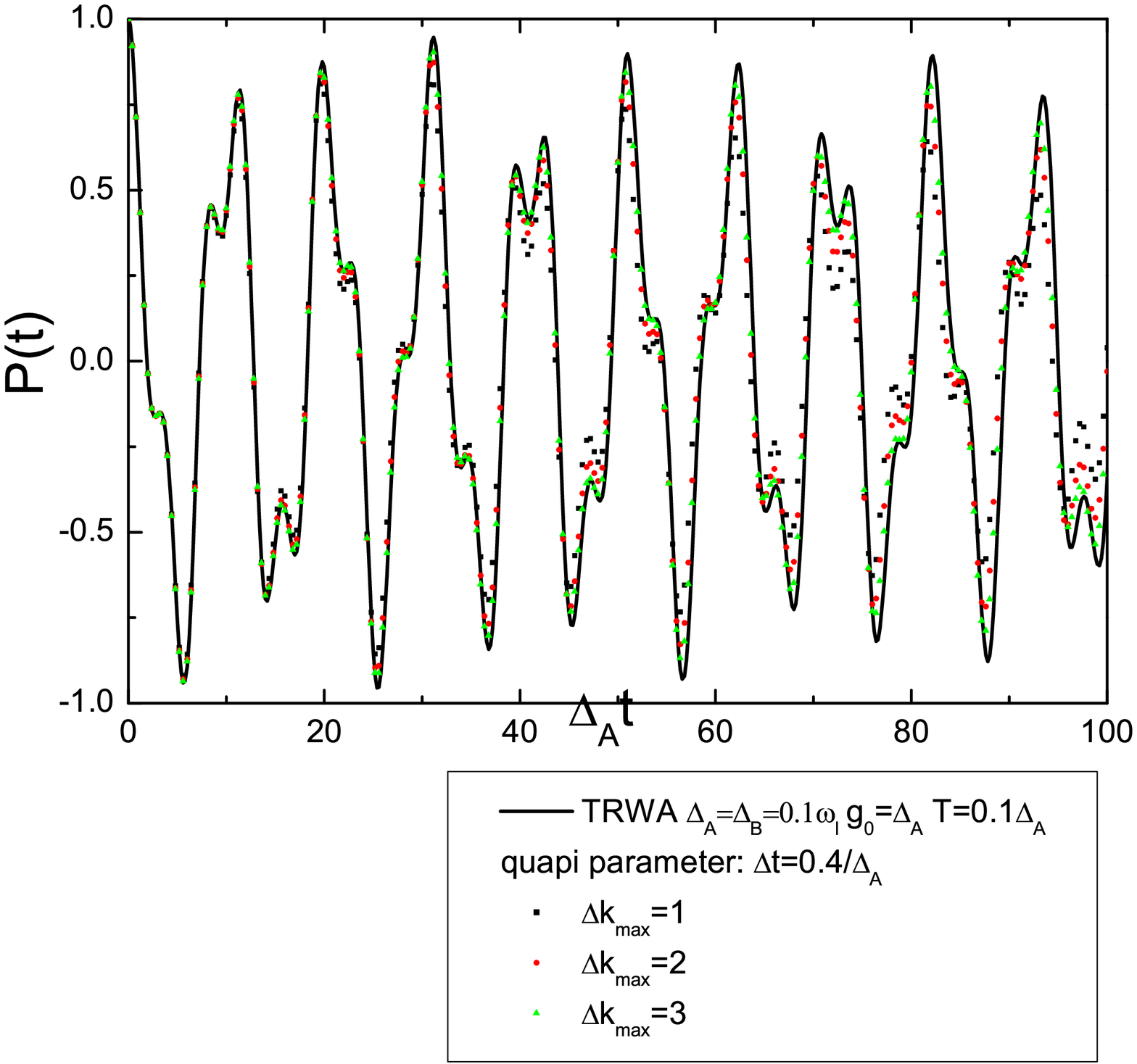}\newline
\end{figure}

\end{document}